\documentclass[sigconf,9pt]{acmart}

\usepackage[english]{babel}
\usepackage{blindtext}
\usepackage{soul}
\usepackage{color, xcolor}

\setcopyright{acmcopyright}

\acmDOI{}

\acmISBN{}

\acmPrice{}

\usepackage{subfigure}
\usepackage{float}
\usepackage{paralist} 
\usepackage[T1]{fontenc}

\begin{document}
\acmYear{2024}\copyrightyear{2024}
\acmConference[EMS '24]{SIGCOMM Workshop on Emerging Multimedia Systems}{August 4--8, 2024}{Sydney, NSW, Australia}
\acmBooktitle{SIGCOMM Workshop on Emerging Multimedia Systems (EMS '24), August 4--8, 2024, Sydney, NSW, Australia}
\acmDOI{10.1145/3672196.3673399}
\acmISBN{979-8-4007-0711-7/24/08}

\title{Towards Real-Time Neural Volumetric Rendering on \\ Mobile Devices: A Measurement Study}


\author{Zhe Wang}
\orcid{1234-5678-9012}
\affiliation{%
  \institution{UM-SJTU Joint Institute\\ Shanghai Jiao Tong University}
}
\email{123369423@sjtu.edu.cn}

\author{Yifei Zhu}
\authornote{Corresponding author}
\orcid{1234-5678-9012}
\affiliation{%
  \institution{UM-SJTU Joint Institute\\ Shanghai Jiao Tong University}
}
\email{yifei.zhu@sjtu.edu.cn}

\renewcommand{\shortauthors}{Z. Wang.et al.}

\begin{abstract}
Neural Radiance Fields (NeRF) is an emerging technique to synthesize 3D objects from 2D images with a wide range of potential applications. However, rendering existing NeRF models is extremely computation intensive, making it challenging to support real-time interaction on mobile devices. 
In this paper, we take the first initiative to examine the state-of-the-art real-time NeRF rendering technique from a system perspective. We first define the entire working pipeline of the NeRF serving system. We then identify possible control knobs that are critical to the system from the communication, computation, and visual performance perspective.
Furthermore, an extensive measurement study is conducted to reveal the effects of these control knobs on system performance. Our measurement results reveal that different control knobs contribute differently towards improving the system performance, with the mesh granularity being the most effective knob and the quantization being the least effective knob. In addition, diverse hardware device settings and network conditions have to be considered to fully unleash the benefit of operating under the appropriate knobs.   
\end{abstract}

\ccsdesc[500]{Computing methodologies~Rendering}
\ccsdesc[500]{Networks~Network performance evaluation}
\ccsdesc[500]{Human-centered computing~Ubiquitous and mobile computing}

\keywords{NeRF, real-time rendering, mobile computing, performance measurement}

\maketitle

\section{Introduction}

Neural Radiance Field (NeRF) is a novel view synthesis method that has gained popularity since its introduction in 2020\cite{mildenhall2020nerf}. It employs a neural network to learn the 3D geometry and appearance of objects and scenes from 2D images, allowing it to render new views with high visual fidelity from any viewpoint. 
Compared with the traditional explicit representations, like mesh and point cloud, the NeRF as an implicit representation can represent complex 3D shapes and scenes with arbitrary topology and generate 3D representations much faster. Since its inception, it has found applications in diverse fields such as virtual and augmented reality\cite{lin2023parallel}, robotics\cite{kuang2023ir}, architecture and interior design\cite{10.1145/3528233.3530718}.


The core concept of NeRF involves volume rendering, which involves the process of sampling hundreds of points to render the appearance color of the pixel that is visible along the viewing ray. Each sampled point must undergo the neural network, making the whole process extremely computation extensive. The original NerF requires 150-200 million network queries per rendered image, taking approximately 30 seconds per frame on an NVIDIA V100\cite{mildenhall2020nerf}. 
This makes it challenging to achieve high-quality real-time performance with NeRF, especially on mobile devices. 


To achieve real-time NeRF-based volume rendering, current strategies mainly involve pre-computing and storing the trained NeRF network’s output before the rendering process so that the actual rendering cost can be significantly reduced. These strategies are also termed as the baked approach in the fields of computer vision and graphics\cite{gao2022nerf}. 
For instance, Mobile-NeRF \cite{chen2023mobilenerf} and its variations \cite{tang2023delicate}\cite{guo2023vmesh} convert the trained Neural Radiance Field into a geometric triangle mesh, then stores the pre-computed features of the object's appearance. Leveraging the stored meshes, and pre-computed features, the mesh-based baking NeRF can achieve at least 200 times faster rendering speed than the traditional NeRF which further leads to a much smoother viewing experience.




Despite widespread interest in real-time neural volumetric rendering within the computer graphics and vision communities, it has not been thoroughly analyzed from a systems perspective. In this paper, we pioneer a system examination of contemporary real-time NeRF rendering efforts. We define the complete workflow of the baked NeRF service system, identify key control knobs affecting communication, computation, and visual performance, and conduct a comprehensive study to assess these knobs' impact on system metrics like file size for transmission cost, visual quality, and rendering speed for computation cost. Based on our findings, we offer insights to optimize the performance of real-time NeRF rendering systems on mobile devices. Our contributions and key insights are summarized as follows:
\begin{itemize}
    \item We conduct the first measurement study of neural rendering on mobile devices from a system perspective.
    \item We assess the importance of different control knobs in existing baked systems, finding that mesh granularity is the most effective knob. 
    \item We reveal that the memory space and other supporting libraries have a significant impact on the effectiveness of these control knobs. These two factors restrict the baked-NeRF's full application and further improvement potential on mobile devices.
    \item We find that the correlation between the patch size and the textural image size exhibits a nearly linear trend, 
   enabling a straightforward linear profiling model. 

\end{itemize}

\section{Background and Related Work}

\subsection{NeRF Basics}
NeRF combines a fully-connected neural network (MLP) and the volume rendering technique together to reconstruct the whole scene or objects with photo-realistic appearance after learning from different view-points photos. The neural network takes a 5-D vector as its input, consisting of the space coordinate of the viewed point and the viewing ray direction ($\theta$ and $\phi$). The outputs of the MLP are the point's corresponding appearance color and its volume density which is used to calculate the transparency. The model weights are optimized based on certain loss functions, such as minimizing the sum of the squared error between the rendered pixel color and the true pixel color\cite{mildenhall2020nerf}. 


For the inference part, users only need to provide the NeRF network his/her view direction and the viewed object's positions in the camera world. Then, each pixel of the final rendering image will have a corresponding viewing ray
and sampling will be applied along the viewing ray. For each sampled point, the NeRF will execute the MLP to get the corresponding color. By accumulating the color with the volume rendering formula, the user can get the final color of the corresponding pixel. By repeating the procedure for every pixel, the new view-point image will be rendered.

\subsection{Real-time NeRF Rendering}


Considering the significant potential in interactive applications, real-time NeRF has attracted wide attention. Current real-time NeRF rendering works can be divided into two categories: precomputing and storing the results of MLPs, i.e. ``baked'' approach, and the one that does not involve such a step, i.e. ``un-baked'' approach\cite{gao2022nerf}. 
The un-baked NeRFs try to learn a separate scene, e.g., sparse voxels\cite{liu2020neural}, instead of dealing with the original one and fastening the sampling procedure for each ray by reducing the number of points or accelerating the access to each point, e.g., via a hash table \cite{muller2022instant}. 


Compared with unbaked mobile-feasible work, the baked methods have less requirement for training data.
The unbaked approach needs 100 times more data than the original one\cite{cao2023real} while the data for the Mobile-NeRF, a baked approach, remains the same. When being deployed in real systems, transmission these extra images also introduces significant communication costs. 
Due to these limitations of the unbaked approaches, in this paper, we mainly focus on examining baked approaches.

\subsection{A Mobile NeRF Service Pipeline}

\begin{figure}[!tbp]
  \centering
  \includegraphics[height = 3.8cm, width=\linewidth]{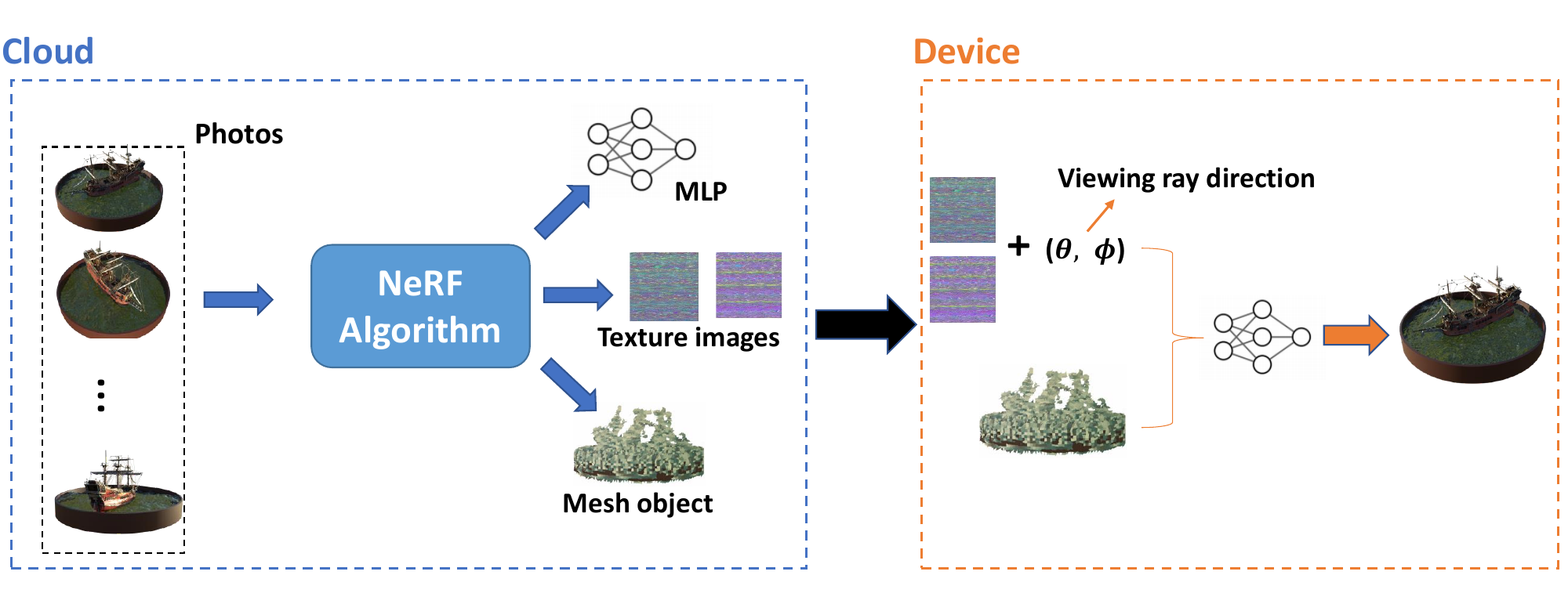}
  \caption{End-to-end pipeline of baked-NeRF serving.}
  \label{pipeline}

\end{figure}

The whole pipeline of a practical baked NeRF serving system can be divided into two parts. One is network training and the other one is rendering. 
The end-to-end pipeline of a bake NeRF model serving is shown in 
Figure~\ref{pipeline}.  Considering the computing requirement for model training, we first have a cloud server that collects photos from users and trains a neural network for 3D representation following certain NeRF algorithms. We use Mobile-NeRF as the case algorithm in this paper because it is a representative baked work that achieves real-time rendering on mobile devices and is filled with diverse pre-computing types to explore. To achieve real-time rendering on mobile devices, this baked NeRF algorithm generates MLP and pre-computes the texture images and the mesh object for later use. 
When users want to consume this 3D object or scene,  they can download all the training products of the corresponding object to their mobile devices through communication networks. Given the viewing direction, the received MLP will output the color based on the textural images. The color will then be combined with the mesh object to get the final output image.

\section{METHODOLOGY}
The existing NeRF service pipeline has three major components from the systems perspective: the mesh object, the textual image, and the MLP. For these three major components, we identify the following control knobs and study their effects on system performance.

\subsection{Mesh Objects}
\label{mesh_obj}

During the baked training process, a grid mesh which consists of {\itshape P} $\times P \times P$ voxels is generated to cover the whole object. Then for each voxel, a vertex inside it will be allocated. By connecting the four adjacent voxels' vertices we can have a quad face (a quad face can be made up of two triangles). By back-propagating the loss, the positions of the vertices can also be optimized to optimally reconstruct an object's 3D polygonal mesh. Therefore, the number of vertices that are used to reconstruct the whole object depends on the parameter {\itshape P}. The larger the {\itshape P} is, the more vertices we can have, and more 3D details can be preserved in the 3D polygonal mesh. On the other hand, the choice of this granularity also may affect training time and transmission cost. Intuitively, the larger the {\itshape P} is, the  finer the granularity becomes and the resulting mesh objects may be larger.  Therefore, we identify the number of voxels in one dimension, P, as the control knob for the mesh objects, and term it as the \textit{mesh granularity}. 



\subsection{Textural image}
The texture image in a baked NeRF system is used for coloring the mesh object. Each quad face of the mesh object has a corresponding region of {\itshape K} $\times$ {\itshape K} in the texture image for rendering its appearance. We call this region as a patch with size {\itshape K} $\times$ {\itshape K}. The smaller the patch size, the lower the resolution will be for the rendering of each quad face.  The whole process of rendering is actually the same as the traditional mesh rendering. During the rendering process, the quad face will be rasterized into {\itshape K} $\times$ {\itshape K} pixels; For each pixel, the corresponding region of the patch will be used to color it. In other words, if the patch size is larger, the rasterization will also be more refined to show more detail of the quad face. 
Therefore, we identify the \textit{patch size} as the control knob for the textural image.





\subsection{MLP Model}
A crucial component in any NeRF system is the MLP, which is notably compact in mobile NeRF systems, with a size of only about 10.7KB in Mobile-NeRF. Given the significance of MLPs in NeRF, exploring potential enhancements from a systems perspective is essential. To accelerate the computation and reduce memory footprint, common neural network compression methods include pruning, quantization, knowledge distillation, etc\cite{han2015deep}. 
To avoid the necessity of retraining, which is time-consuming for growing interactive applications, in this study, we explore the neural network \textit{quantization level} as the control knob for the MLP component\cite{gholami2022survey}. 
In particular, we apply the post-training quantization technique, which maps the traditional FP64 activation space to a smaller space to further reduce its preparation overhead. Smaller activation space may accelerate inference time, and reduce the model size with the risk of decreasing the accuracy.



\subsection{Measurement Setup}
\textbf{Dataset and Devices}
We test the performance of volumetric rendering with all the eight synthetic 360-degree objects provided by \cite{mildenhall2020nerf}. 
In this paper, we use two different mobile devices to do the measurements study. They are (1) iPhone13: 6-core CPU with 3.23 \& 2.01GHz, 4GB RAM and Apple-designed four-core GPU; (2) Pixel4: eight-core CPU with 2.84 \& 2.42 \& 1.78GHz, 6GB RAM and Adreno 640 GPU. 

\textbf{Metrics} We measure the system performance in terms of communication cost, visual quality, and smoothness. The corresponding metrics are as follows:
\begin{itemize}
    \item \textbf{File size}
    The communication cost in the whole pipeline comes from the downloading as users have to download all the auxiliary files for rendering. Thus, we can use the file size to quantify the communication cost.
    \item \textbf{Visual quality}
    We use the PSNR (Peak Signal-to-Noise Ratio) and the SSIM (Structural Similarity Index Measure) as the visual quality metric.
    PSNR is calculated by comparing the original image with the reconstructed one, and finding the ratio between their maximum possible pixel value and their mean squared error. PSNR is usually expressed in decibels (dB), where higher values mean better quality.
    The SSIM compares the luminance, the contrast, and the structure of two images \cite{wang2004image}. The relation between the SSIM and the similarity is the same as PSNR.
    
    \item \textbf{Frames per second (FPS)}
 Given a specific hardware device, FPS reflects the computation cost during the rendering process. Higher FPS leads to smoother playback during the interaction.
\end{itemize}

\textbf{Configuration space for control knobs}
The default patch size is 17$\times$17; The default mesh granularity is 128; the default quantization level is FP64. To examine the effect of all control knobs, we divide the mesh from 32 to 128 equally into 7 stages, \{32, 48, 64, 80, 96, 112, 128\}.
We set the upper bound as 128 to match the default value of the existing real-time NeRF rendering studies\cite{chen2023mobilenerf} \cite{hedman2021baking}. The lower bound is set to be 32 because setting it lower already makes it difficult to recognize typical objects.
The patch sizes {\itshape K} $\times$ {\itshape K} we select are 9$\times$ 9, 17$\times$17 and 33$\times$33 \footnote{Note that the real region of the texture patch are 8$\times$8, 16$\times$16 and 32$\times$32 respectively because there exists a half-a-pixel boundary padding.}. 
We quantize the models into different digital bits: 8, 4, and 2, so that the parameter can be presented by FP32, FP16, and FP8 respectively. 


\textbf{Implementation}
We use  the webGL as the rendering engine on browsers for interaction. For iPhone 13, we use Safari, and for Pixel4, we use Chrome.  Because the generated MLP is in the configuration of Jax which doesn't have a mature library of model optimization and has immutable properties, we first transfer the MLP into the pytorch and realize the quantization based on the quantization library of tensorflow.  We set the reconstructed object to rotate at a fixed speed (7.5 seconds to rotate 360 degrees) to generate a uniform viewing behavior for a fair comparison. We then measure the average FPS of the whole viewing process.

\section{RESULTS AND ANALYSIS}
We measure the effect of the identified control knobs on communication, visual quality, and computation.

\subsection{Effect of Mesh Granularity}


    
    
    
    

\begin{figure}[!tbp]
    \centering
    \subfigure[]{\includegraphics[width=4cm]{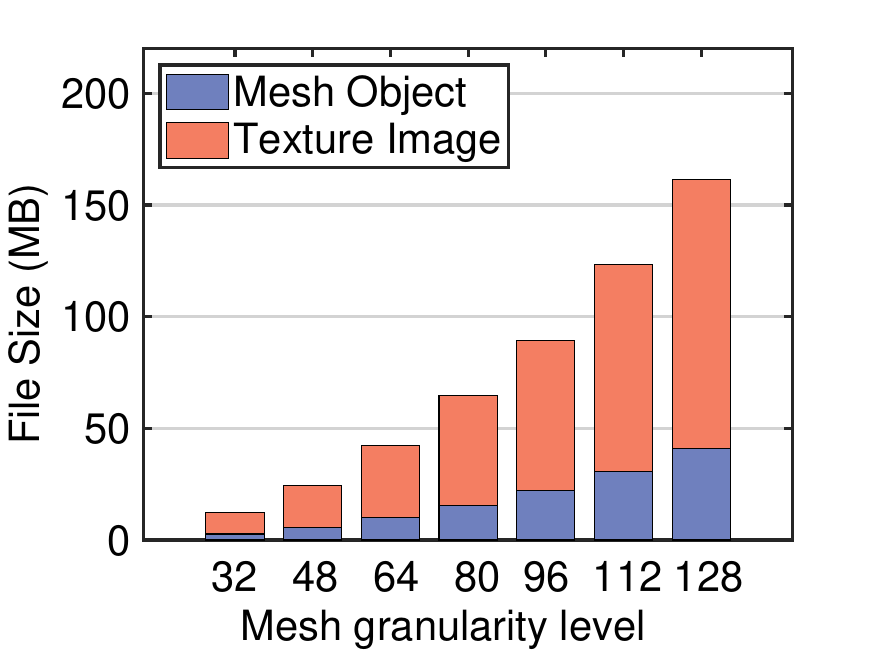}\label{size_granularity}}
    \subfigure[]{\includegraphics[width=4cm]{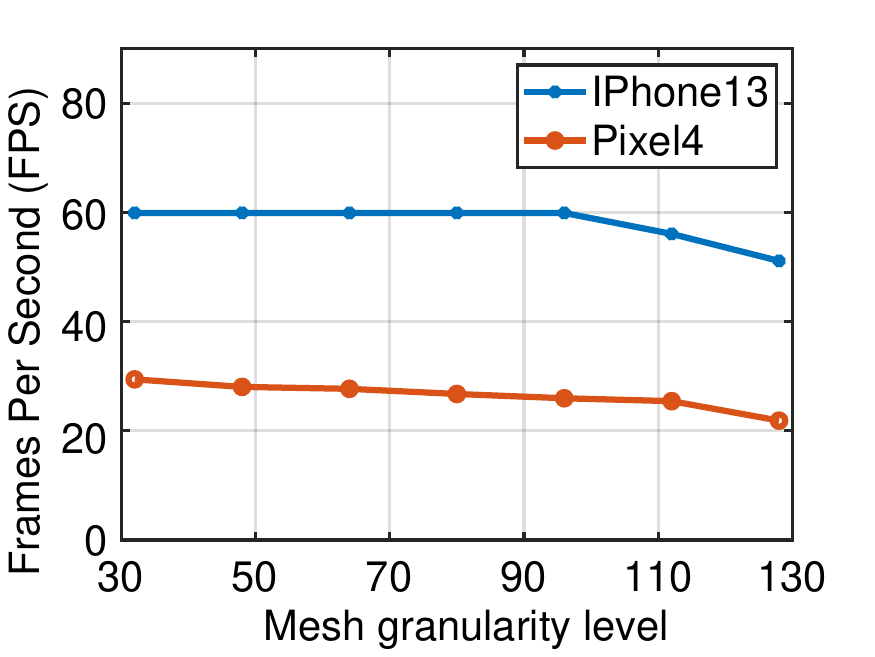}\label{fps_granularity}}
    \caption{File size and rendering speed under different mesh granularity}
    \label{communication_FPS_gran}
    \end{figure}
        
\textbf{File size} The file sizes under different granularity are presented in Figure~\ref{size_granularity}. As can be seen, mesh granularity can affect both the size of the mesh objects and the size of the texture image. With the increase of the mesh granularity, both the size of the mesh objects and the textural images increases. 
The total file size of texture images ranges from 78.1\% of the total file size to 74.7\% of the total file size. The proportion of the texture images stays around 75\% of the total file size under different granularity levels. 
The size of MLPs stays at about 10.7 KB when the mesh granularity level changes from 32 to 128. Since their value is small compared with the rest two types, we omit plotting them out. 


    \begin{figure}[!tbp]
    \centering
    \subfigure[]{\includegraphics[width=4cm]{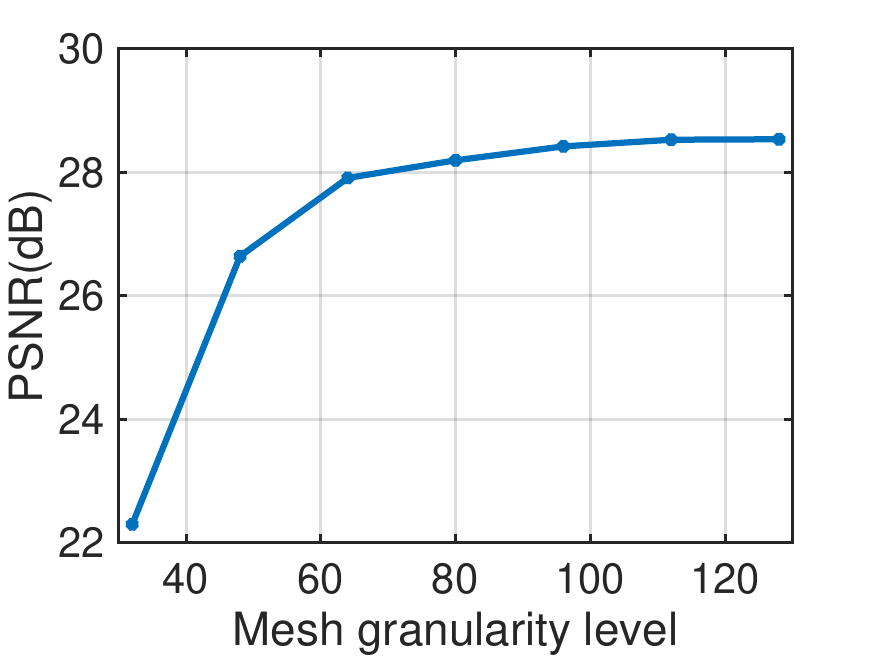}}
    \subfigure[]{\includegraphics[width=4cm]{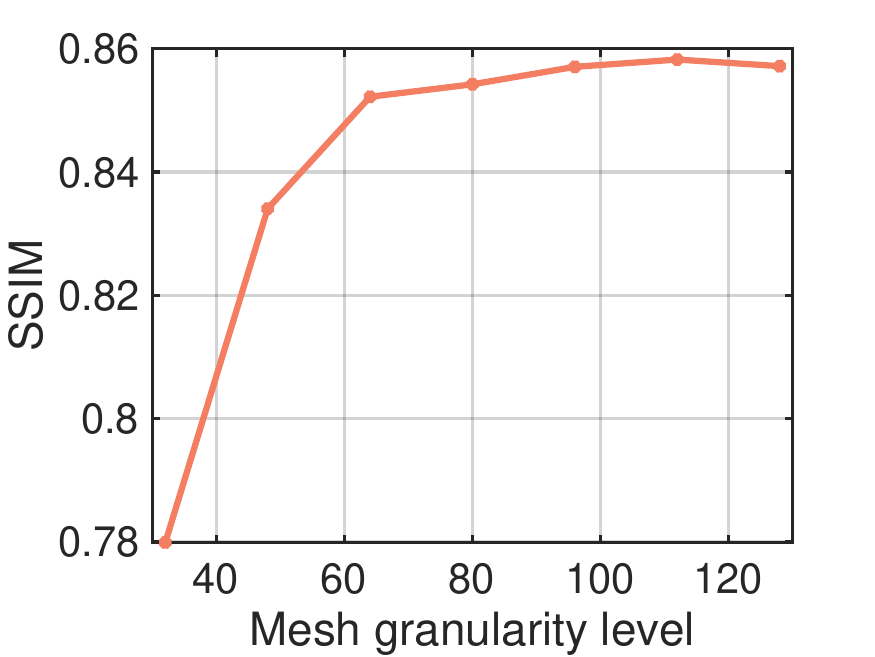}}
    \caption{Visual quality, PSNR and SSIM, under different mesh granularity}
    \label{quality_gran}

    \end{figure}
    
    \begin{figure*}[!tbp]
    \centering
    \subfigure[The file size under different patch sizes with granularity 80 and 128]{\includegraphics[width = 0.3\linewidth]{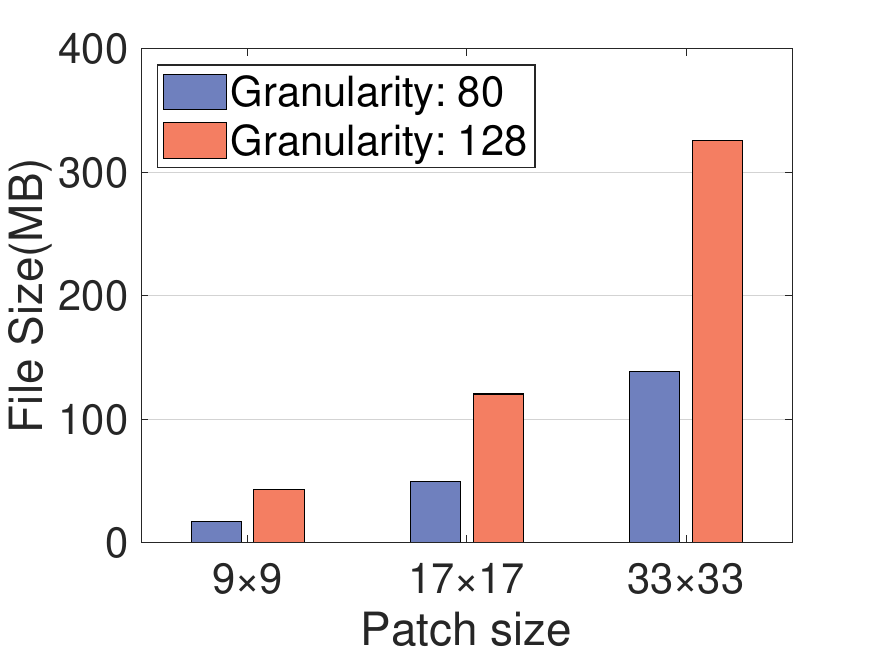} \label{size_text}}
        \subfigure[Convergence FPS under different patch size with granularity 80 ]{\includegraphics[width=0.3\linewidth]{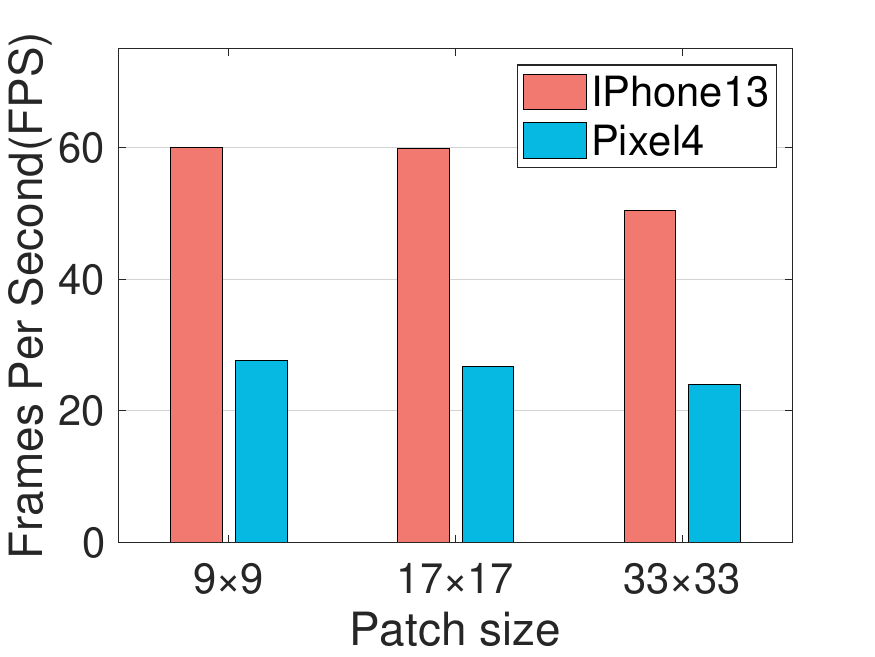} \label{fps_text_80}}
    \subfigure[Convergence FPS under different patch size with granularity 128]{\includegraphics[ width=0.3\linewidth]{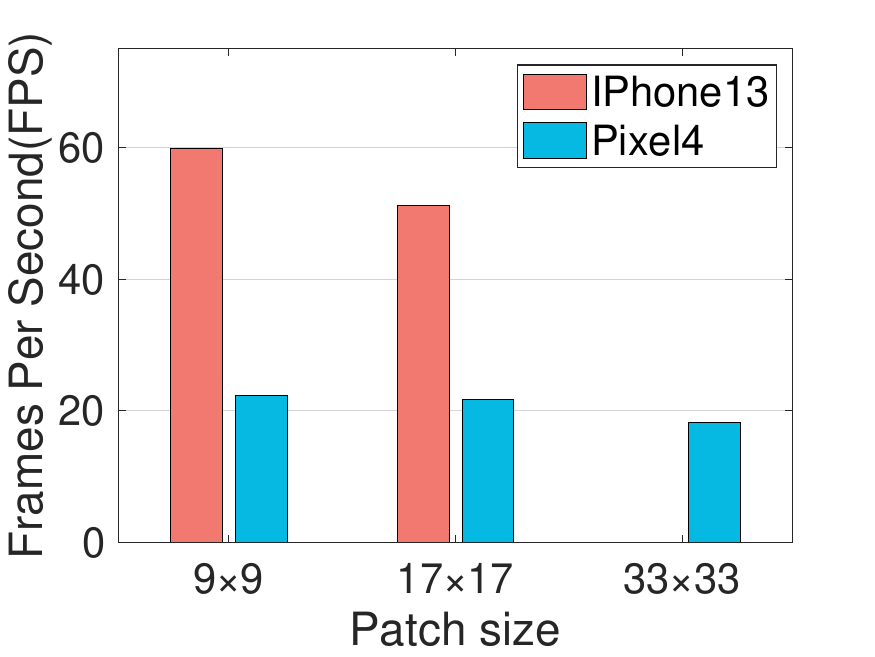} \label{fps_text_128}}
    \caption{File sizes and rendering speed under different texture image patch size and mesh granularity level}
    \label{texture_measure}
    \end{figure*}

\textbf{Visual quality} We present the effect of mesh granularity on visual quality in Figure \ref{quality_gran}.
Under the default patch size (17 $\times$ 17), PSNR is about 22.294 under the granularity of 32 and it can reach 28.533 when the mesh granularity level reaches the finest level, 128. PSNR value only reduce 2.1\% when the mesh granularity changes from 128 to 64. While the file size can reduce 75.6\% for the mesh object and 73.1\% for the texture image when the mesh granularity changes from 128 to 64. Similarly, SSIM is about 0.780 under the lowest granularity level of 32 and can further reach 0.857 when the granularity level reaches the finest level, 128. It only reduces 0.5\% when the mesh granularity changes from 128 to 64. 

    
\textbf{The FPS performance}  The convergence FPS rates on iPhone13 and Pixel4 under different mesh granularity are shown in Figure~\ref{fps_granularity}. For the case of iPhone 13, the converged FPS rate remains at 60 across mesh granularity ranging from 32 to 96 and then starts to drop while in the case of Pixel 4, the convergence FPS rate continuously drops with the granularity increasing. The difference in FPS is due to the different computing capabilities of the devices. 

From the measurement results, we can derive the following observations: (1) Compared with the MLP, the auxiliary files to facilitate real-time rendering bring non-negligible  communication costs. Optimization is urgently needed to reduce this cost to make the whole system practical. (2) Adjusting mesh granularity can generate huge benefits in reducing the file size with little sacrifice on visual quality.



    

\subsection{Effect of Patch Size}

    \begin{figure}[!tbp]
    \centering
    \subfigure[]{\includegraphics[height=3.5cm, width=4.1cm]{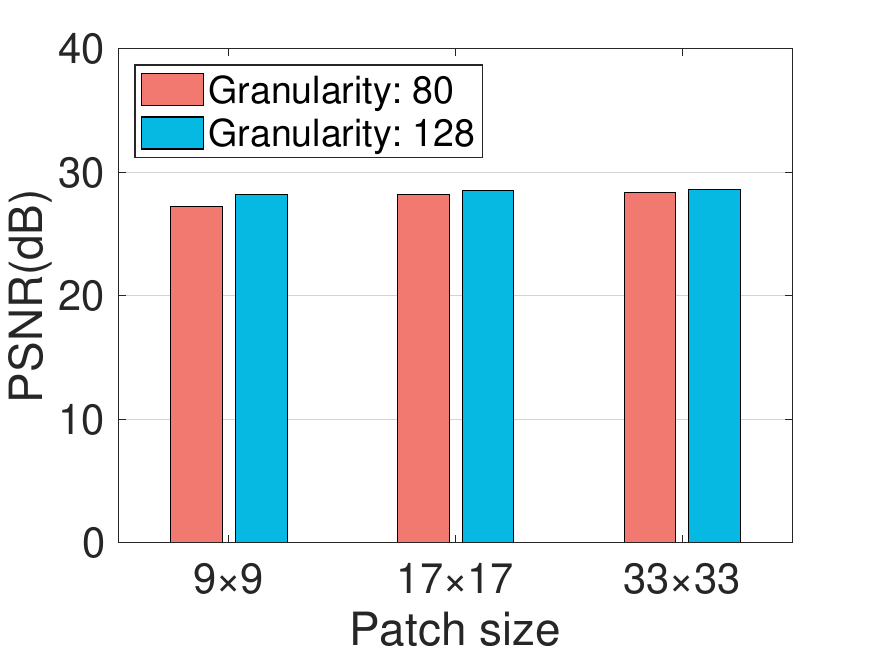}}
    \subfigure[]{\includegraphics[height=3.5cm, width=4.1cm]{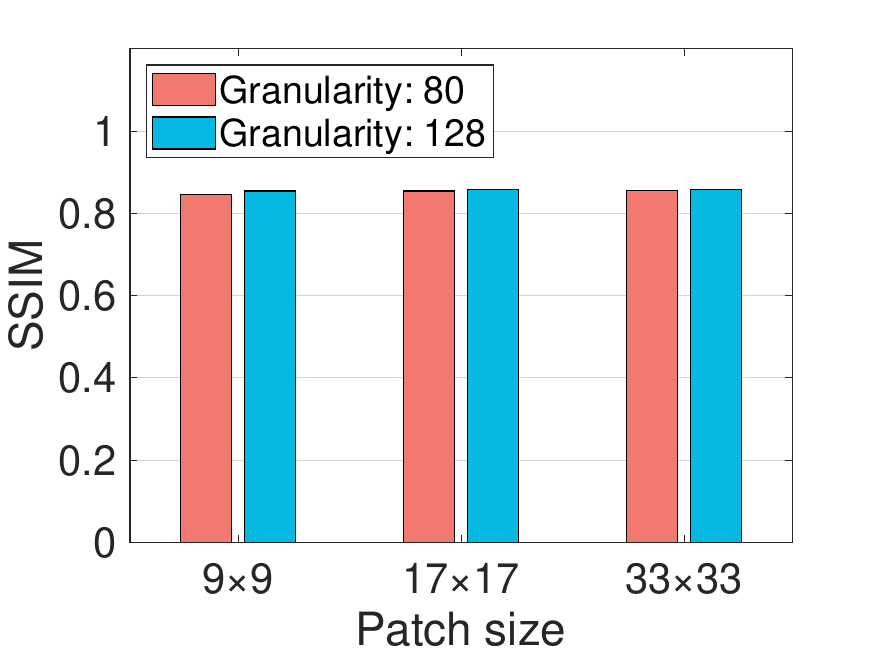}}
    \caption{Visual quality under different texture image patch size with granularity 80 and 128}
    \label{quality_patch}
    \end{figure}
    
    \textbf{File size} 
    Since the patch size only determines the size of the textural image, we report the resulting textural image size in Figure. \ref{size_text}. From the results, the textural image size grows almost linearly with the increase of the patch size.  In both cases (granularity with 80 and granularity with 128), the patch is growing at the rate of twice while the texture sizes are growing at the rate about 2.7.

    \textbf{Visual quality} The visual quality measurement results are in the Figure~\ref{quality_patch}. From the size and quality measurements results, it's quite clear that although the increase of patch size does improve the rendering quality, the improvement is quite small no matter under which granularity. The {\itshape K} changes from 9 to 17 and 17 to 33, both sizes of products become three times as much as before, but the increase on PSNR values changes from almost 1 to only about 0.2 while the one on SSIM values changes from about 0.1 to 0.001. 

    \textbf{FPS performance} Figure~\ref{fps_text_80}  shows the rendering speed result under granularity 80 while Figure \ref{fps_text_128} contains the results under granularity 128.  
    In Figure~\ref{fps_text_80}, the convergence FPS of iPhone13 is 60 under the patch size of 9$\times$9 and 17$\times$17. Then it will drop about 15.8\% to 50.5 under 33$\times$33 patch size. The FPS changing tendency of Pixel4 under the granularity is quite similar. As the change of FPS from patch size 9$\times$9 to 17$\times$17 is only about 0.88. But when changing from 17$\times$17 to 33$\times$33, the reduction is 2.82. The Pixel4's tendency on patch size remains under the 128 granularity level as shown in Figure~\ref{fps_text_128}. The FPS dropping from 17$\times$17 to 33$\times$33 patch size (3.36) is much larger than the one from 9$\times$9 to 17$\times$17 (0.66). For iPhone 13, the FPS drops about 8.87 under 128 granularity from patch size 9$\times$9 to 17$\times$17. 
    Following the same reason as we introduced in the mesh granularity part, the FPS performance remains the same under 9$\times$9 and 17$\times$17 with granularity 80 because of the hardware limitation. It is also worth mentioning that the convergence FPS becomes 0 on iPhone13 when the patch size is 33$\times$33, as shown in figure\ref{fps_text_128}. 
This primarily results from the file sizes of the generated products being too large for iPhone 13 to handle, making the whole rendering process crash. 

From the measurement results, we can derive the following key observations: 
(1) The relationship between the patch size and the textural image size is almost linear, making a simple linear predictor possible for predicting the effect of patch size on textural size. 
(2)  Though current baked approaches could potentially lead to real-time rendering performance, the extra burden on memory may make the whole scheme infeasible to deploy.

\subsection{Effect of Model Quantization}
    
    

\begin{figure}[!tbp]
    \centering
    \subfigure[The MLP size of the original MLP and the quantized MLPs on iPhone13]{\includegraphics[height=3.7cm,width=4.1cm]{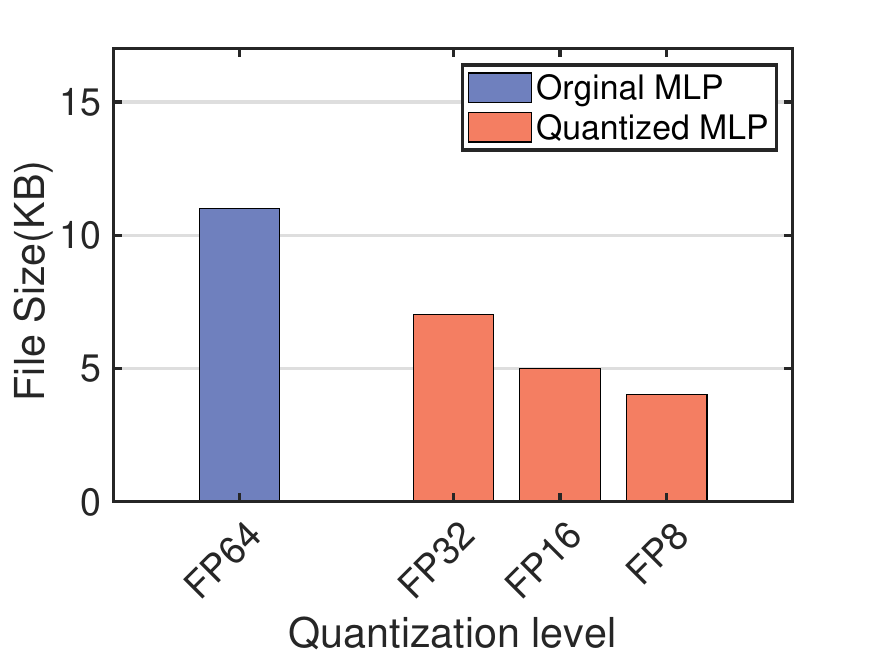}\label{fps_q_s}}
    \subfigure[The time cost of different quantized MLPs]{\includegraphics[height=3.7cm,width=4.2cm]{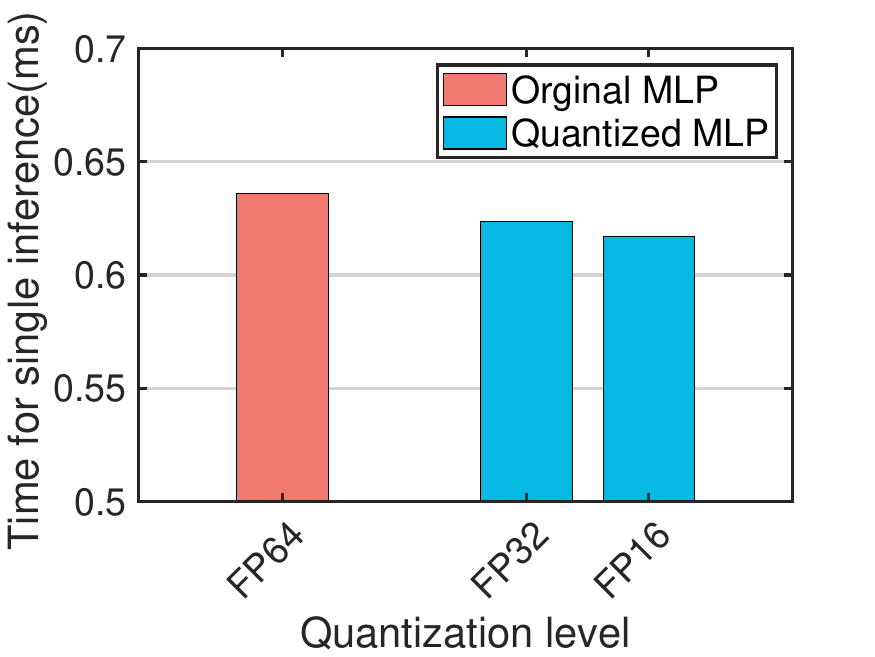}\label{fps_q_i}}
    \caption{The measurement results of quantized MLPs }
    \label{FPS_q}
    \end{figure}

\textbf{File size}
    Based on Figure~\ref{fps_q_s},
    quantization effectively reduces the size of the MLP. Initially, the MLP with FP64 is about 10.7KB. With the three different quantization strategies: FP32, FP16, and FP8, the size can be reduced to 6.84KB, 4.42KB and 3.41KB which are 64\%, 41\%, and 32\% of the original MLP respectively, with almost no visual quality loss.
    
    \textbf{Visual quality} 
    The effect of quantization on the visual quality is shown in the Table~\ref{Quality_q}. 
    Each row of the table presents the result of a comparison between the specific quantization strategy with the original MLP.
    In this table, the PSNR value continuously decreases as the quantization precision gets lower. When the quantization level is FP32, the PSNR can exceed 100; it reduces to 50.79 when we use FP8 to represent the parameters with only 2 decimal bits. Although the value has dropped about 50\%, it is still above 50 which means even with FP8 data type that only has about two digital bits, the rendering results are still extremely similar to the results from the original MLP. All the SSIM values in the three quantization strategies can reach 0.999 which is quite close to the upper bound 1 of the SSIM value. Both PSNR and SSIM all indicate that the quantization can be enforced without sacrificing too much visual quality. 

    \begin{table}[t]
    \caption{The rendering visual quality comparison between the original MLP and the quantized MLPs }
    \label{table1}
    \centering  
    \setlength{\tabcolsep}{3mm}{
    \begin{tabular}{c|c|c}
    \hline
    Degree of quantization & PSNR (dB) & SSIM \\
    \hline
    FP32 & 106.15 & 0.99999 \\
    \hline
    FP16 & 73.18 & 0.99996 \\
    \hline
    FP8 & 50.79 & 0.99905 \\
    \hline
    \end{tabular}}
    \label{Quality_q}
    \end{table}
    
\textbf{Computation performance} 
We record the time for single round of MLP inference to accurately determine the effect of quantization on MLP inference speed in Figure~\ref{fps_q_i}. The
performance of FP8 data type is not recorded because it is not supported in Pytorch. Under the original FP64 activation space, the time for the MLP to do one forward inference is about 0.636 ms. After quantization to FP32 activation space with 8 digital bits, the time cost can be reduced about 0.013 ms to 0.623 ms. Compared with the FP16 activation space with 4 digital bits, the time cost reduces by about 0.019ms to 0.617 ms per forward inference. The quantization method accelerates the inference speed of the MLP. However, the effect is not significant for single-round inference.

From the measurement results, we can have the following observations: (1) Our examined post-quantization methods can reduce file size and accelerate processing time, but the effect is relatively small compared with other control knobs. (2) Current mobile devices should be further enhanced to support different quantization techniques to fully unleash the power of quantization in computation time acceleration. Without a mature optimization that supports other data type operations, the performance can not get corresponding improvements. 
    
\section{Discussion}

 \subsection{Dominant Control Knobs}
    
    
The roles that mesh granularity, patch size, and quantization play are different. The mesh object stores the whole scene/object's geometry structure while the texture images are used for the appearance color. It's not practical to make the two parameters as large as possible for the entire service system. Although this way does improve the final rendering practical, it can also put a significant burden on the system. The excessive cost of both transmission and computation can easily exceed the mobile device's capability. Thus, it's necessary to verify which baked data is more important to the overall system performance. Our measurements study showed that the mesh granularity level has much more significant impacts on the communication cost, the rendering quality as well as the computation cost. Although the texture image can improve the rendering quality, it makes more cost on the communication and computing side at the same time. 

    
    

    

\subsection{Extra Requirements on Hardware}
 The overall performance of the NeRF inference can be greatly affected by the devices. The most obvious evidence is on the FPS performance comparison between iPhone 13 and Pixel4. In the Mobile-NeRF, under the same mesh granularity 80 and patch size 17, the rendering speed in FPS has a significant difference between the two devices. This difference is mainly caused by the computation capability of each CPU. In the case of Pixel4 and iPhone13, the Pixel4 is clearly limited by its computing power, with all its FPS performances below 30, while the iPhone13 is limited by the memory size. Therefore, when the system serves the clients, these factors should all be taken into consideration.
    

    

\subsection{The Importance of Network Condition}
The network may impact the rendering process in scenarios involving multiple mesh objects. Because the mobile device renders each group of products (a mesh and its corresponding feature images) as they are received. If another group of objects cannot be downloaded or loaded promptly, the rendering process may suffer significant delays, severely degrading the user experience with the baked NeRF service. For example, in Mobile-NeRF's rendering of a ship floating on the sea, users might initially see only the sea and, depending on their network conditions, wait for the ship to appear after the rest of the data is downloaded. Thus, it is crucial to ensure either the completeness of the baked result or that high-priority content blocks are streamed first.


\section{Conclusion and future work}
In this paper, we take the first initiative to examine the mesh-based real-time NeRF volume rendering framework from a system perspective. We identify three major control knobs in a real-time NeRF rendering system that can affect its communication, computation, and storage cost. We measure the benefits they could bring to these system dimensions on real-world mobile devices. 
Based on the measurement results, we identify several key observations: (1) mesh granularity is the dominant control knob. (2) Users' hardware device settings enforce non-negligible constraints for our examined services. (3) Quantization is less effective in improving the performance than the rest two knobs. Furthermore, adopting this technique is still further constrained by the existing mobile software libraries and chipset support. 

Beyond our current focus on mesh-based NeRF volume rendering, alternative NeRF algorithms utilize varied architectures or leverage different deep learning frameworks to synthesize static 3D objects, such as the previously mentioned unbaked NeRF. Future studies can be done to identify promising control knobs and examine the effects of these control knobs on these structural variations. Additionally, we plan to extend our evaluations to NeRF algorithms tailored for dynamic scenes or video content, ensuring a thorough understanding of their applicability and performance across diverse scenarios.

\section*{Acknowledgment}
This work is supported by the National Natural Science Foundation of China (Grant No. 62302292) and the Fundamental Research Funds for the Central Universities.  

\bibliographystyle{ACM-Reference-Format}
\bibliography{reference}

\end{document}